\documentclass[11pt]{article}
\usepackage{moriond,psfig}

\def\Journal#1#2#3#4{{#1} {\bf #2}, #3 (#4)}

\def\be{\begin{equation}}
\def\ee{\end{equation}}
\def\bea{\begin{eqnarray}}
\def\eea{\end{eqnarray}}

\def\HI{\hbox{H~$\scriptstyle\rm I\ $}}

\def\HeII{\hbox{He~$\scriptstyle\rm II\ $}}

\def\kmsmpc{\,{\rm km\,s^{-1}\,Mpc^{-1}}}

\def\Lya{Ly$\alpha\ $}

\def\spose#1{\hbox to 0pt{#1\hss}}
\def\lta{\mathrel{\spose{\lower 3pt\hbox{$\mathchar"218$}}
     \raise 2.0pt\hbox{$\mathchar"13C$}}}
\def\gta{\mathrel{\spose{\lower 3pt\hbox{$\mathchar"218$}}
     \raise 2.0pt\hbox{$\mathchar"13E$}}}

\begin{document}
\vspace*{4cm}
\title{MODELLING THE UV/X-RAY COSMIC BACKGROUND WITH {\it CUBA}}

\author{F. HAARDT$^{1}$ \& P. MADAU$^{2}$}

\address{~$^{1}$Dipartimento di Scienze, Universit\`a dell'Insubria, Como, Italy 
~$^{2}$Department of Astronomy \& Astrophysics, University of California, Santa Cruz, USA}

\maketitle\abstracts{
In this paper, I will describe the features of the numerical code CUBA, aimed at 
the solution of the radiative transfer equation in a cosmological context. 
CUBA will be soon available for public use at the URL http://pitto.mib.infn.it/~haardt/, 
allowing for several user-supplied input parameters, such as favourite cosmology, luminosity functions, 
Type II object evolution, stellar spectra, and many others. 
I will also present some new results of the UV/X--ray cosmic background as
produced by the observed populations of QSOs and star forming galaxies, updating and
extending our previous works \cite{hm96} \cite{mhr}.
The background evolution is complemented with a number of derived quantities such as the ionization and thermal 
state of the IGM, the \HeII opacity, the \HI and \HeII ionization rates, and the \HI, \HeII and
Compton heating rates.} 

\section{Introduction}

The existence of an ubiquitous medium which contains most of the baryons present in the Universe at any epoch is
    predicted by current cosmological theories of structure formation. The absence of a Gunn-Peterson absorption
    feature shortward the Lyman alpha emission line in the spectra of background quasars results in stringent limits
    on the fraction of gas still present in neutral form in the all pervading inter-galactic medium (IGM). It is now
    known that the IGM is highly ionized, at least up to the highest redshifts we observe quasars. The recent
    observation of the quasar J104433.04-0125022 with redshift z=5.8 \cite{fan} sets the current upper limit of
    the reionization epoch.

    The issue concerning the exact epoch of reionization, and the nature of the ionizing sources, has been, in recent
    years, the target of a large number of studies by leading scientific groups worldwide. Two different approaches
    emerged from such studies. One is mainly theoretical, and it is based on the theory of hierarchical clustering of
    structures. It has been realized that the chemistry of molecular hydrogen played a crucial role in the formation of
    the first luminous objects. Roto-vibrational line emission from H2 could have been the first effective coolant for
    baryons within collapsing dark matter (DM) halos. The minimum temperature for H$_2$ cooling to be effective sets
    a precise epoch of the formation of the first stars, which turns out to be at $z\simeq 30$. 
The successive evolution of such
    primordial stellar populations within dwarf proto-galaxies is usually followed by means of parametric models,
    which have the advantage of parameterizing our ignorance for what concerns the IMF, the escape fraction of
    Lyman limit photons, the supernova feedback, etc \cite{bc} \cite{go} \cite{g00} \cite{hl98}.
    On the other hand, these parameters are necessarily tuned to present day values. AGN activity has
    been also postulated in the earliest forming galaxies, due to the possible presence of a supermassive black hole
    lurking in the center of most of the dark matter halos \cite{l93} \cite{hmen} \cite{hl98} \cite{hl99}. 
All these studies set the epoch of reionization at $z\simeq 10-12$.

A second, more phenomenological, approach is instead focused on the observed part of the Universe, i.e., on the
    observed known sources of UV/X-ray radiation. The basic idea is to compute the propagation of a radiation
    background model due to known sources, and to test possible observable effects on the IGM. Such approach
    resides on observed quantities such as luminosity functions, redshift evolutions, spectral energy distributions,
    absorbing cloud distribution, etc, and is clearly complementary, not alternative, to the first methodology discussed
    above. 
In this context, the nature of the ionizing and heating sources has been longly debated in the last few years 
\cite{sg} \cite{m92} \cite{hm96} \cite{gs} \cite{mhr} \cite{h01}.
An obvious source of
    high energy radiation such as bright quasars has been matched, in terms of possible energy input in the IGM, by
    the large number of high redshift young star forming galaxies (YSFGs) observed in deep fields \cite{s01}. 
On one hand it
    has been demonstrated that quasar alone did not produce enough ionizing photons to explain the observed upper
    limit of the reionization epoch \cite{mhr}. On the other hand, it has been proved to be quite
    difficult to asses the precise role of YSFGs in the formation of a UV background, because of the poorly known
    escape fraction of Lyman limit photons \cite{hu}. 

Despite the mentioned difficulties, such second approach is a valuable tool to study the thermal history of the IGM.
Back in 1994, we started to develop a numerical code aimed at the solution of the radiative transfer problem 
in an expanding Universe. The code, named CUBA (after Cosmic Ultraviolet BAckground), 
has been described in Haardt \& Madau \cite{hm96}, 
and our results have been extesively used either to model the IGM in large 
comsological simulations, and in interpreting observations \cite{gw} \cite{d97} \cite{t98} \cite{s99} \cite{s00}. 

We are currently updating our modelling of the background, considering the most recent determinations of QSO
    luminosity function and redshift evolution, stellar SEDs from Bruzual \& Charlot synthesis code, Lambda
    cosmology, extending the analysis in the X-ray regime. The latter point is of particular importance, 
as the the latest observations by Chandra and XMM-Newton have resolved a large fraction of the 0.5-10 keV X-ray
    background (XRB) into discrete sources, mainly AGNs. Synthesis models for the XRB require that a numerous
    population of type II AGN does exist at moderate large redshifts (0-2). Such absorbed AGN population, detectable
    in the hard X-ray band, should give a non negligible contribution in the IR, due to the thermal re-emission of the
    X-ray absorbing molecular torus. Moreover, as pointed out by Madau \& Efstathiou \cite{me}, Compton heating of the 
IGM due to the XRB could be as important as photoionization heating due to the UVB.  
We will describe in detail our new results in two forthcoming papers \cite{hm01} \cite{mh01}. 
In the present contribute, instead, we mainly focus on the code itself, on its features and performances, as CUBA will 
be soon available to the community.

\section{CUBA}

CUBA comes actually as a package, containing, together with CUBA itself, several other codes which, in most of the cases, 
can be run independently, once that the proper input files have been created. 
CUBA solves the radiative transfer equation, and produces the 
file containing the specific intensity as a function of wavelength and redshift. 
As byproducts, several other quantities are sorted out. 
Some of the newly created files can then be used as input files for the ancillary codes, 
whose output are X-ray number counts, fitting 
parameters to the heating and ionization rates, temperature of the IGM, etc.  
As mentioned above, one of the new feature of CUBA is the extension of the calculation in the X-ray band.
The package contains also a simple, completely independent Montecarlo code which allows the user to build 
up a synthetic Type II spectrum template from the scratch. Such template serves then as an input file for CUBA, 
to calculate an XRB model.

CUBA models the propagation of different combinations of AGN-like and stellar-like
ionizing radiation through the IGM, including QSO absorption-line systems as
sources and sinks of ultraviolet radiation, and QSOs and 
YSFGs as sources of radiation. Other, more exotic sources, can be, in principle, easly accounted for. 

The 1-D equation of radiative
transfer in an expanding Universe is solved by means of an iteration scheme given a specified convergence 
criterium. 
The specific intensity
is computed in a grid of 432 reference wavelengths, covering (with different resolution) roughly
the range from 0.1 mm to 10 MeV, starting from $z\simeq 9$ to
z=0, in 50 redhifts equally spaced in log. 
The details of the microphysics included can be found in Haardt \& Madau \cite{hm96}. The most important
property of CUBA resides in its speed. 
Thanks to the functional relation between the redshift and
wavelength grids, which is very effective in reducing the number of integrations required
at each timestep,
the transfer equation is typically solved in the $432\times 50$ $(\lambda, z)$ points,
in about 90 seconds of CPU by a standard PC equipped with a 1 GHz processor.

The input source function, i.e., the specific emissivity of QSOs and YSFGs, must be specified.
For the QSO population, the specific emissivity vs. redshift is calculated from the 
luminosity function, redshift evolution, and rest-frame SED template.
For what concerns the contribute of YSFGs to the source function, the methodology 
is slightly different. We started from the measured rest-frame UV luminosity at
1500 \AA, assumed to arise from a young stellar population \cite{m96}. 
The prescription for de-reddening this value has been the
subject of a long standing debate. In the present contest, however, the key issue is
not the estimate of the intrinsic star formation rate from fluxes at 1500 \AA~, but
rather is to estimate the ionizing flux at 912 \AA~ from the observed
1500 \AA~ flux. This means one has to take a given stellar SED template, and estimate
 the extra absorption (with respect to the observed 1500 \AA~ photons) the 912 \AA~ photons 
may suffer before escaping from the galaxy. The key parameter is then the so called {\it escape 
fraction} $f_{\rm esc}$ (of Lyman limit photons), though such term is somewhat misleading. 
A straight defintion of the escape fraction is $f_{\rm esc}=L_{\rm esc}(912)/L_{\rm int}(912)$, 
i.e., the ratio between escaped and produced Lyman limit photons. An estimate of $f_{\rm esc}$ from the 
observed flux at 1500 \AA~ then requires to asses the role of internal \HI {\it and} dust absorption.
In CUBA, $f_{\rm esc}$ is defined as 
\begin{equation}
f_{\rm esc}={[L(912)/L(1500)]_{\rm esc}\over [L(912)/L(1500)]_{\rm int}},
\end{equation}
i.e., it is a differential (i.e., normalization free), rather than absolute, measure of the \HI absorption. 
With such choice, the dust issue becomes irrelevant for our purposes. 

The second main ingredient in the radiative transfer equation is the absorption term, 
which is an integral of the column density and redshift distrbution of 
\Lya clouds, and Lyman limit systems. As discussed by Fardal. et al \cite{f98}, simple single power-law fits to the 
column distribution do not provide a good description of the data, due to a (possible) defincency of clouds with 
$N_{\rm HI}\simeq 10^{16}$ cm$^{-2}$. However, in the evolution of the UVB, what counts is the effective opacity, which 
is an integral over the column distribution. Because of the average slope of the cloud distribution ($\simeq 1.5$), the 
effective opacity is dominated by clouds with $\tau_{\rm eff}\simeq 1$, so that the precise fit to the lowest and highest 
column densities is, basically, irrelevant. Indeed, a single power-law approximation to the 
column distribution can be proved to give results within few percent to those obtained with more detailed 
fitting formulae. 
In CUBA we implemented a $N_{\rm HI}^{-1.5}$ description of the column distribution. This permits 
to perform analytical integrations, speeding up the convergence.

Along with CUBA, five rotuines are used to compute quantities related to the UVXRB, e.g., the \HeII opacity, 
the temperature of the IGM (for various overdenisties), the photo-heating, Compton heating and ionization rates with 
useful anlytical fits to them, the log N-log S in the soft and hard X-ray bands.

\subsection{X-ray background}

CUBA is set up as to build up models for the XRB, assuming that the background is due to a mixture of Type I and Type II 
sources. The X-ray evolution of Type II objects, which is the basic free-parameters of the XRB, is computed using 
the procedure discussed in Madau et al. \cite{m94}. Montecarlo simulations of Type II spectra are performed 
for a range of column densities of the obscuring torus. The "average" Type II spectrum si then made up by a convolution 
of the absorbed spectra with a column density log normal distribution.

We will include in the downloadable files a small archive of Montecarlo outputs, so that no prior work before running 
CUBA is needed. The archive will contain absorbed Type I spectra simulated for "standard" values of the input 
parameters. The user will suffice to specify the mean and dispersion of the log normal column distribution. 
In any case, we plan to include also the Montecarlo code itself, so that XRB models could be computed from the scratch. 

\section{Outputs}

In this section I would like to show some of the outputs of CUBA. Given the source function (see Fig. 1) and the 
absorption term, among the quantities computed are  
the background specific intensity (Fig. 2), 
heating and photoionization rates and analytical fits to them (Fig. 3), \HeII opacity (Fig. 4), 
IGM temperature (Fig. 5), AGN number counts in 
the soft and hard X-ray bands (Fig. 6). 
As already mentioned, the required CPU time is about 90 seconds on a PC.
The plots shown refer to a specific input model, and a complete 
description and discussion will appear in two forthcoming 
papers \cite{hm01} \cite{mh01}. Here I just review the main differences with respect to Haardt \& Madau \cite{hm96}. 
The main new ingredient is the inclusion of galaxies as sources of ionizing radiation. The star 
formation rate vs. redshift as computed in Madau et al.\cite{m96} is used to normalize the 1500 \AA~ rest frame flux of 
the stellar radiation, assuming $f_{\rm esc}=0.1$. 
The YSFG template SED is computed from Bruzual \& Charlot libraries, assuming a 
metallicity of 0.2 solar, Slapeter IMF, and an age of 0.5 Gyrs. 
Concerning the QSOs, we adopted the recent determination of the optical LF in a $\Lambda$CDM cosmology given
by Boyle et al. \cite{b00}, resulting in a local emissivity in the B band of $4.8\times 10^32$ erg
Gpc$^{-3}$ s$^{-1}$ Hz$^{-1}$, while the redshift evolution of $L_{\star}$ objects follows the
analytical fit described in Madau et al. \cite{mhr}. 
The spectral slope of QSOs shortward 1050 \AA~ is now steeper, $\alpha=1.8$, after Laor et al. \cite{l97} and 
Zheng et al. \cite{z98}. 
Finally, a flat Universe cosmology is adopted, with non--zero cosmoligical costant ($\Omega_{\Lambda}=0.7$,
$\Omega_{\rm M}=0.3$), and $H_0= 65\,\kmsmpc$.
For what concerns the X-rays, we have adopted a redshift dependent Type II/Type I ratio. A detailed discussion of 
the XRB model will be presented in Madau \& Haardt \cite{mh01}. 

The velocity of a single computation makes CUBA ideal for surveying large 
areas of the parameter space in a short time.
As an example, in the following I will compare the model described above, 
with two other models computed for different valuex of $f_{\rm esc}$. 
In the first model 
$f_{\rm esc}=0$, i.e., only QSOs contribuite to the background. This can be considered a sort of "lower limit" model. 
In the second model, $f_{\rm esc}=0.05$, i.e., the escape fraction is consistent with the local value. In the 
third model we set $f_{\rm esc}=0.05$ for $z<3$, and $f_{\rm esc}=0.5$ for $z>3$, i.e., we assume that high redshift 
YSFGs have an escape fraction larger than what observed locally, consistent with the value estimated by Steidel 
et al. \cite{s01}. In Fig. 7 the emissivity 
at the (rest frame) Lyman limit is shown vs. redshift for QSOs an YSFGs separately, for the three models. 
In Fig. 8 the resulting \HeII/\HI ratio expected in the IGM is plotted. 
Note how the much softer stellar-like spectrum boosts the value of the predicted 
\HeII/\HI ion ratio. A complete analysis will be presented in Haardt \& Madau \cite{hm01}. 

\section{Conclusions}
I have briefly reviewed the basic charachteristics of the the numerical code
CUBA, aimed at the study of the cosmic evolution of the UV and X-ray backgrounds. 
CUBA is fast, so that it is well suited for quick surveys of large parameter spaces (e.g., the escape fraction 
of Lyman limit photons from YSFGs). 
A (hopefully) user-friendly version of the code will be soon available for the interest researchres.

\begin{figure}
\psfig{figure=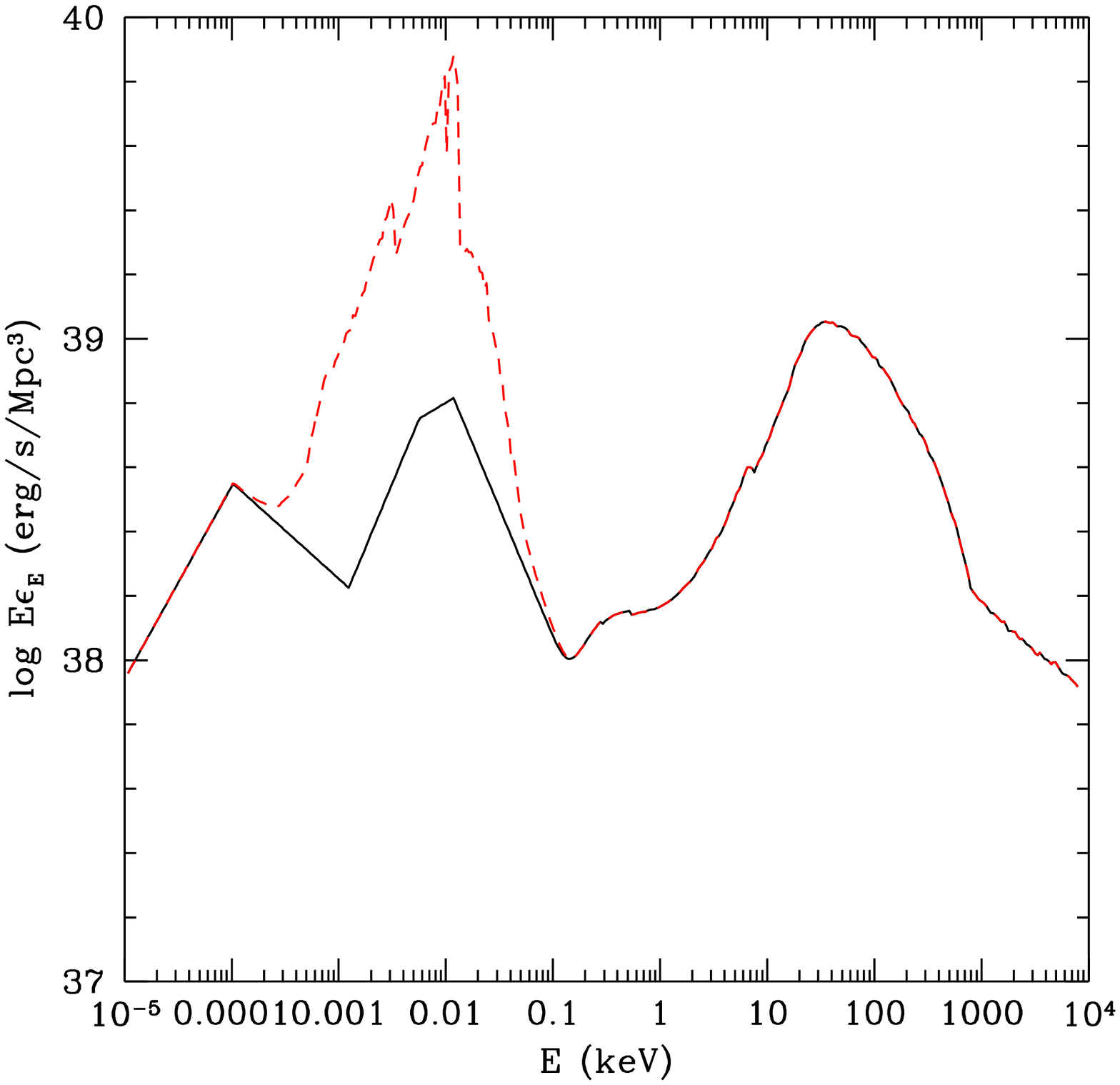,width=1\textwidth,height=0.7\textheight}
\caption{Local emissivity at $z=0$. The red dashed curve is the total emissivity, the black solid line shows 
the contribute of QSOs alone.
}
\end{figure}

\begin{figure}
\psfig{figure=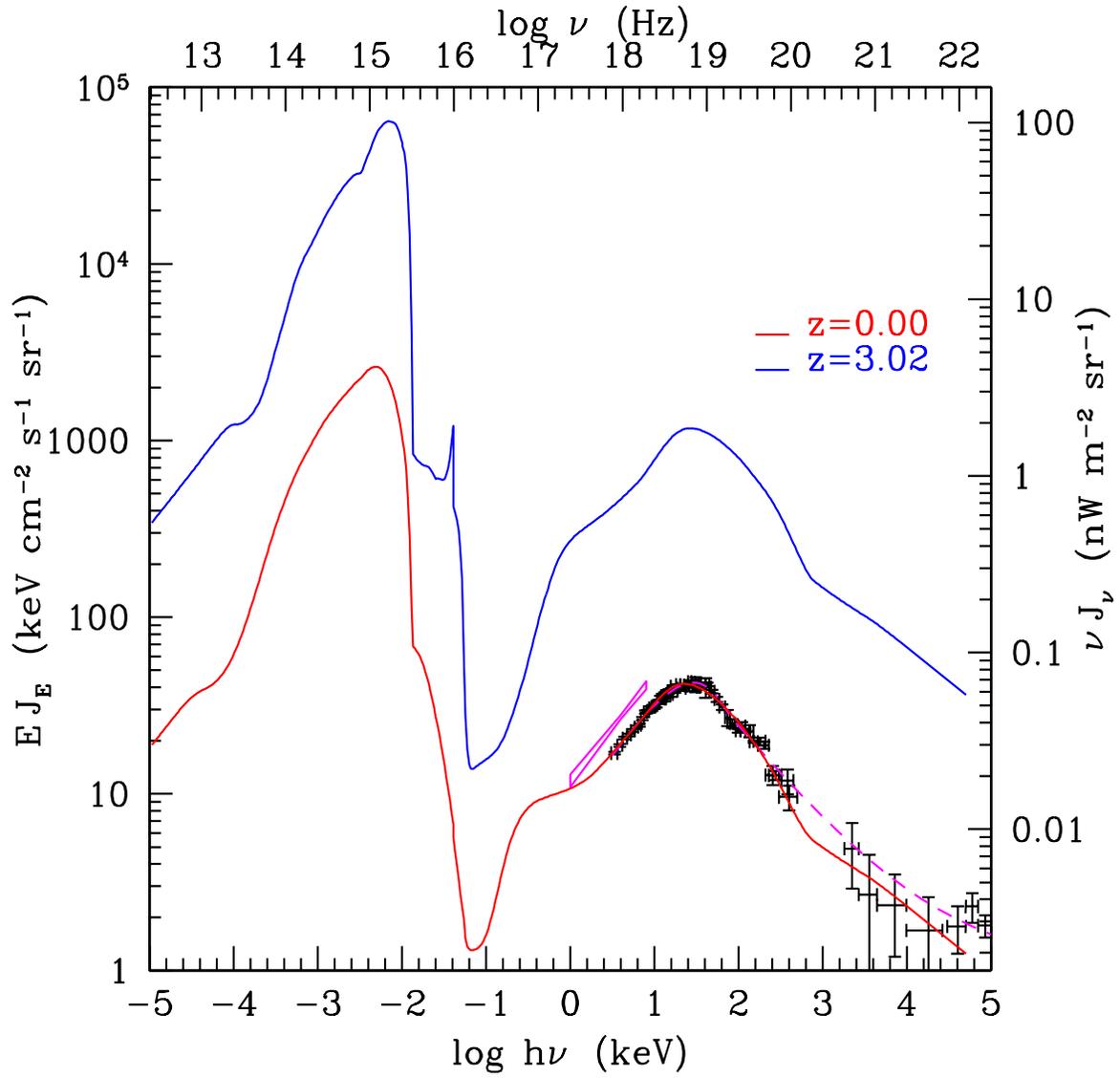,width=1\textwidth,height=0.7\textheight}
\caption{
The cosmic radiation background from IR to $\gamma-$rays, at $z=0$ and $z=3$. 
The X--ray data points at $z=0$ are from HEAO1, while the ebow refers to BeppoSAX data.  
}
\end{figure}

\begin{figure}
\center{
\psfig{figure=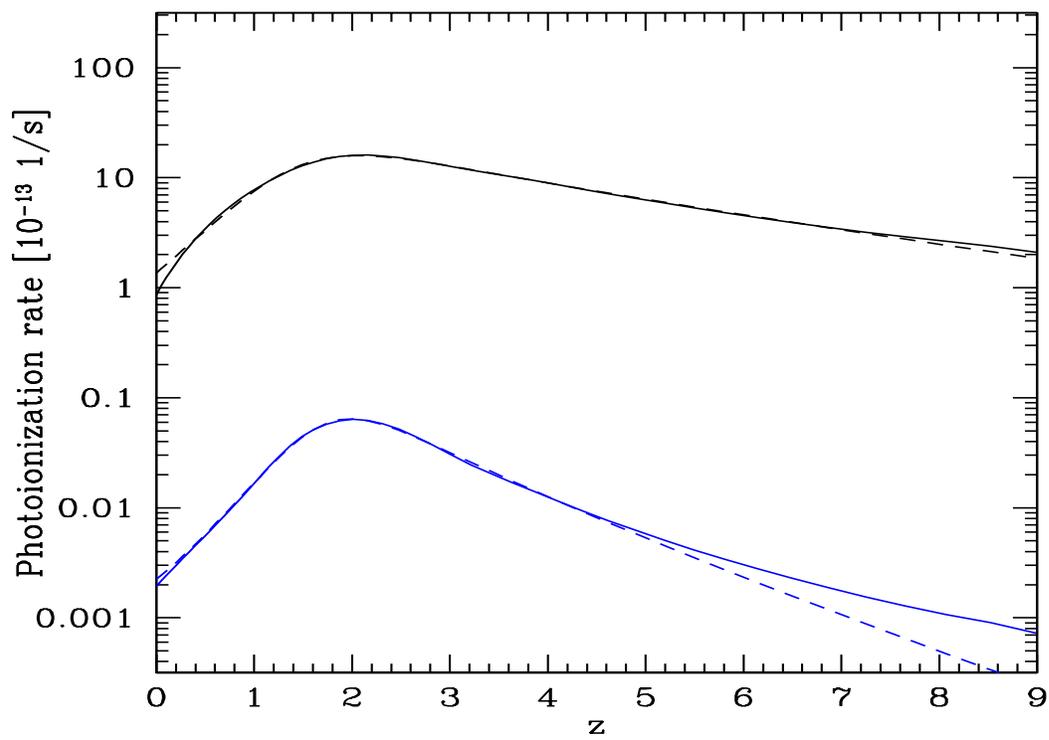,width=1\textwidth,height=0.5\textheight}
\psfig{figure=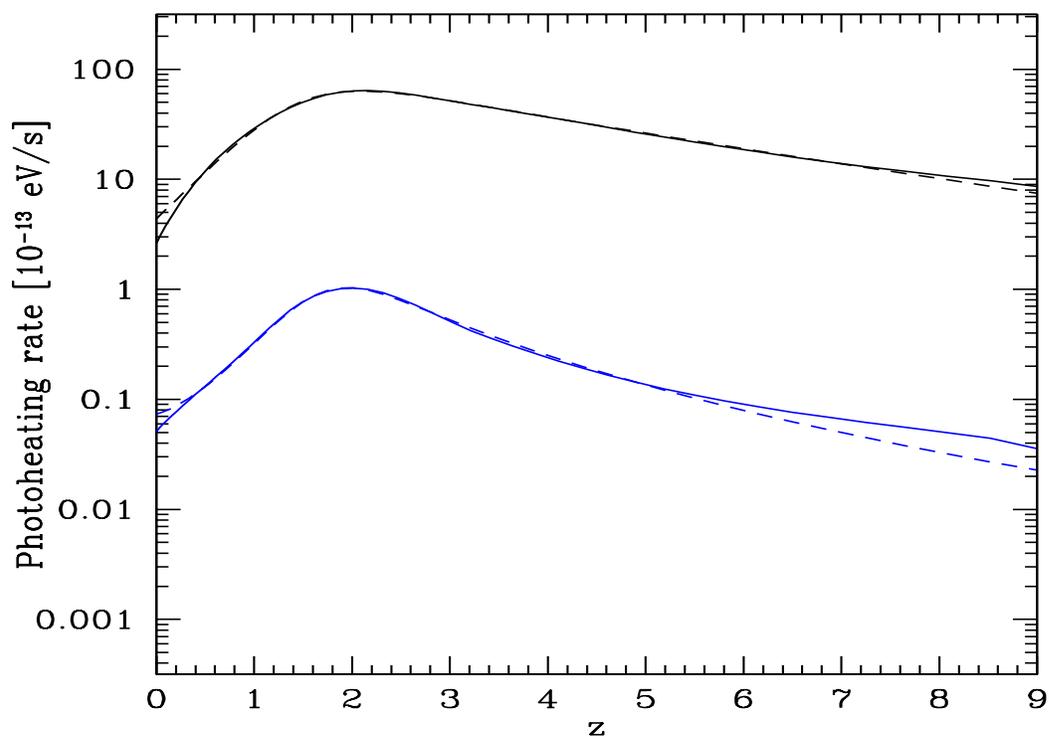,width=1\textwidth,height=0.5\textheight}
}
\caption{Photionization rate (upper panel) and photoheating rate (lower panel) per \HI ion (black lines) and 
\HeII ion (blue lines). Dashed lines are analytical approximations. 
}
\end{figure}

\begin{figure}
\psfig{figure=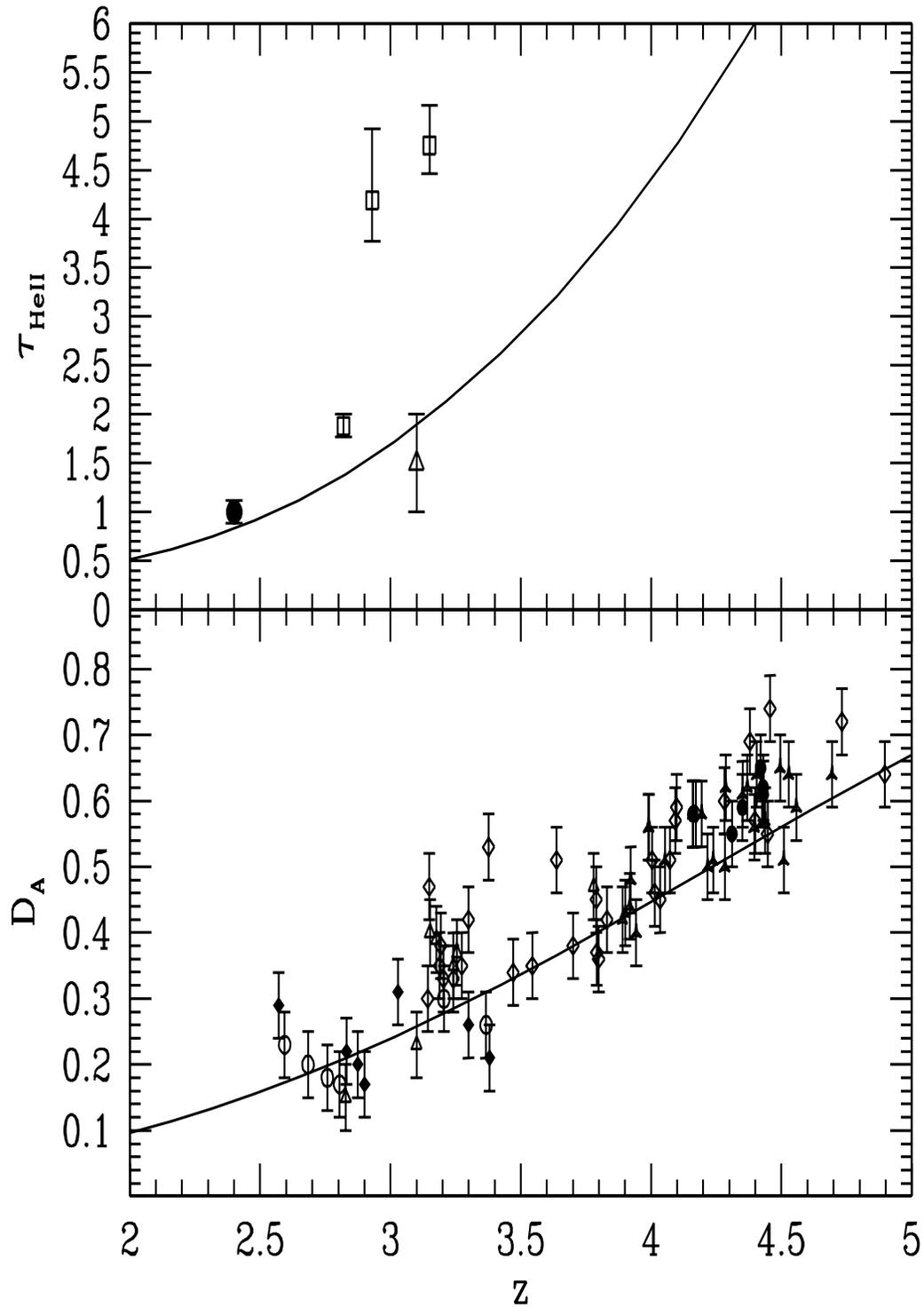,width=1\textwidth,height=1\textheight}
\caption{$D_A$ (lower panel) and $\tau_{\rm HeII}$ (upper panel) vs. z. 
The $D_A$ data are taken from the literature (see Haardt \& Madau$^{1}$  for 
references). The \HeII data are from Davidsen et al.$^{31}$ (filled circles), 
Heap et al.$^{32}$ (empty squares), and Jakobsen et al.$^{33}$ (triangle).}
\end{figure}

\begin{figure}
\psfig{figure=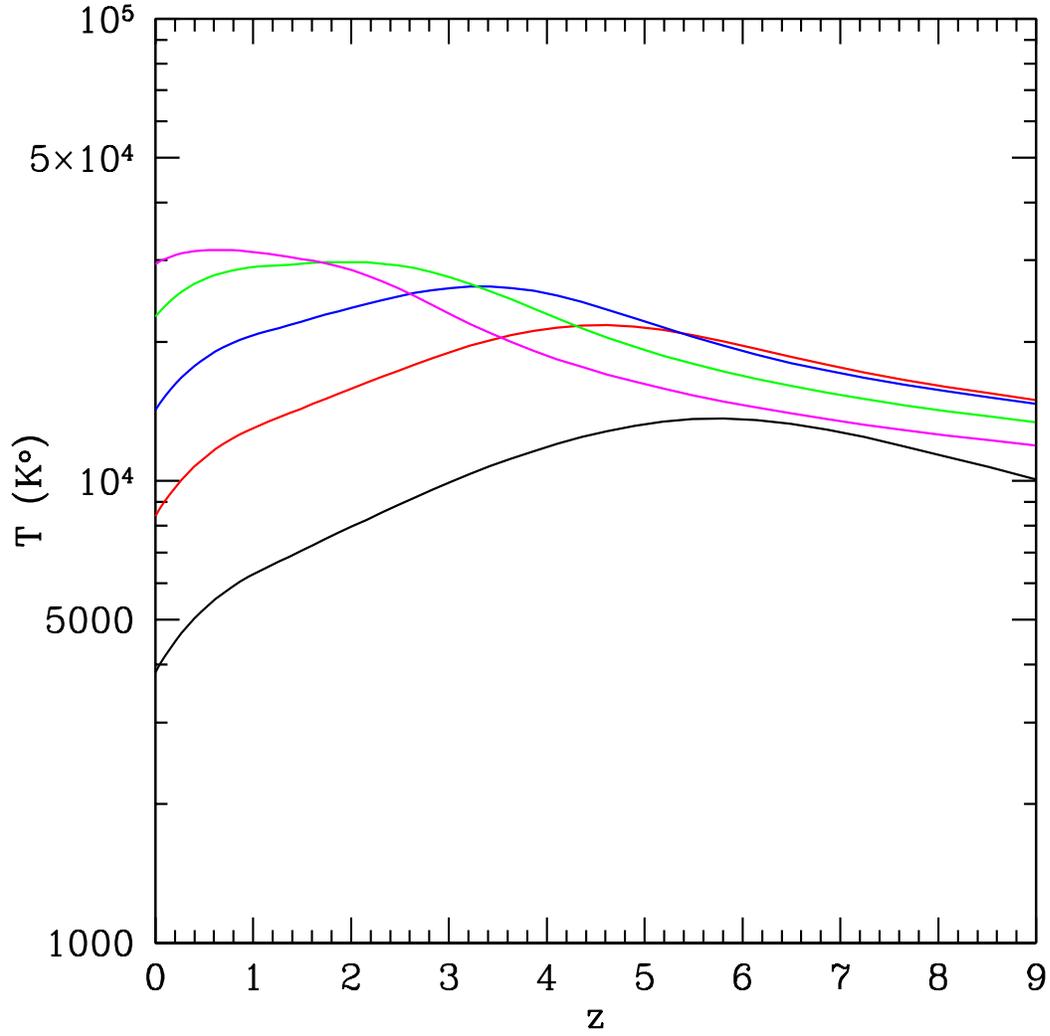,width=1\textwidth,height=0.7\textheight}
\caption{The equilibrium temperature of the IGM for various overdensities ($\delta=1,3,10,30,100$ are the 
black, red, blue, green and purple curve, respectively). 
The temperature is computed assuming thermal equilibrium, by balancing photoionization and Compton heating provided by
the UVB and XRB, with free--bound and free--free radiative losses, collisional excitation losses,
Compton cooling from scattering off the microwave background, and adiabatic cooling due to the Hubble
expansion. 
}
\end{figure}

\begin{figure}
\center{
\psfig{file=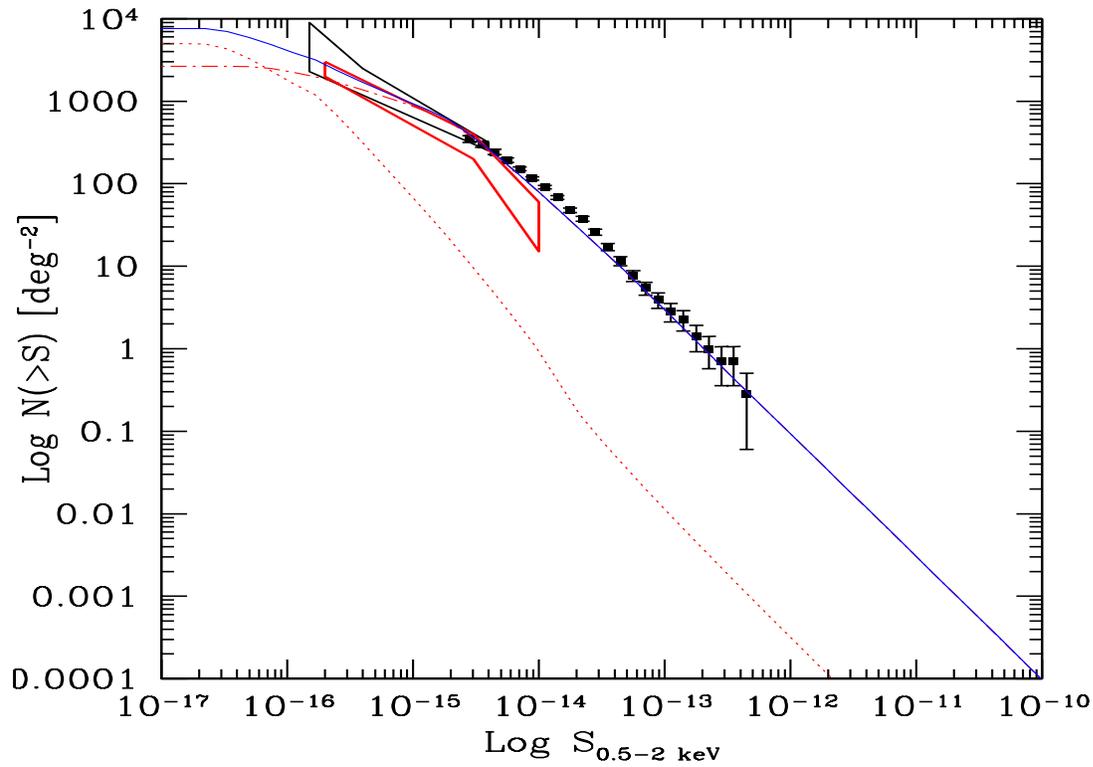,angle=0,width=1\textwidth,height=0.5\textheight}
\psfig{file=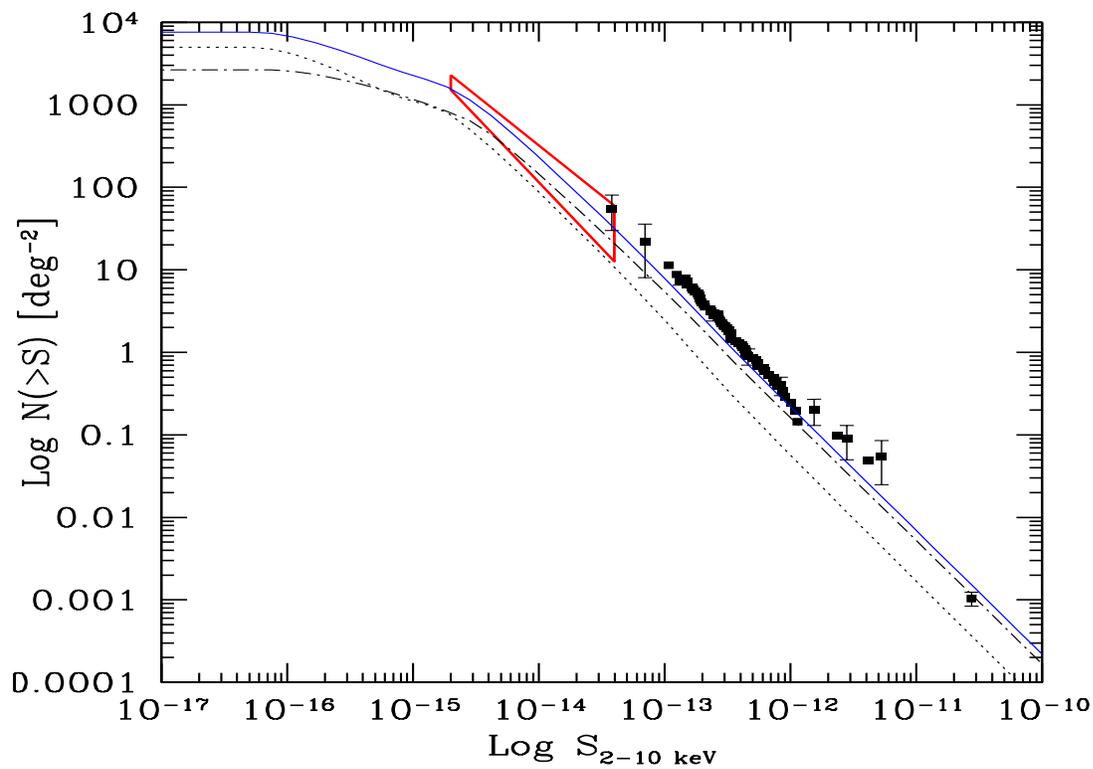,angle=0,width=1\textwidth,height=0.5\textheight}
}
\caption{Soft band (upper panel) and hard band (lower panel) AGN number counts. The data shown are 
from various spaceborne missions, such as EXOSAT, ASCA, GINGA, ROSAT and Chandra. Type I object (dashed lines) and 
Type II objects (dotted lines) are also shown separately.} 
\end{figure}

\begin{figure}
\psfig{figure=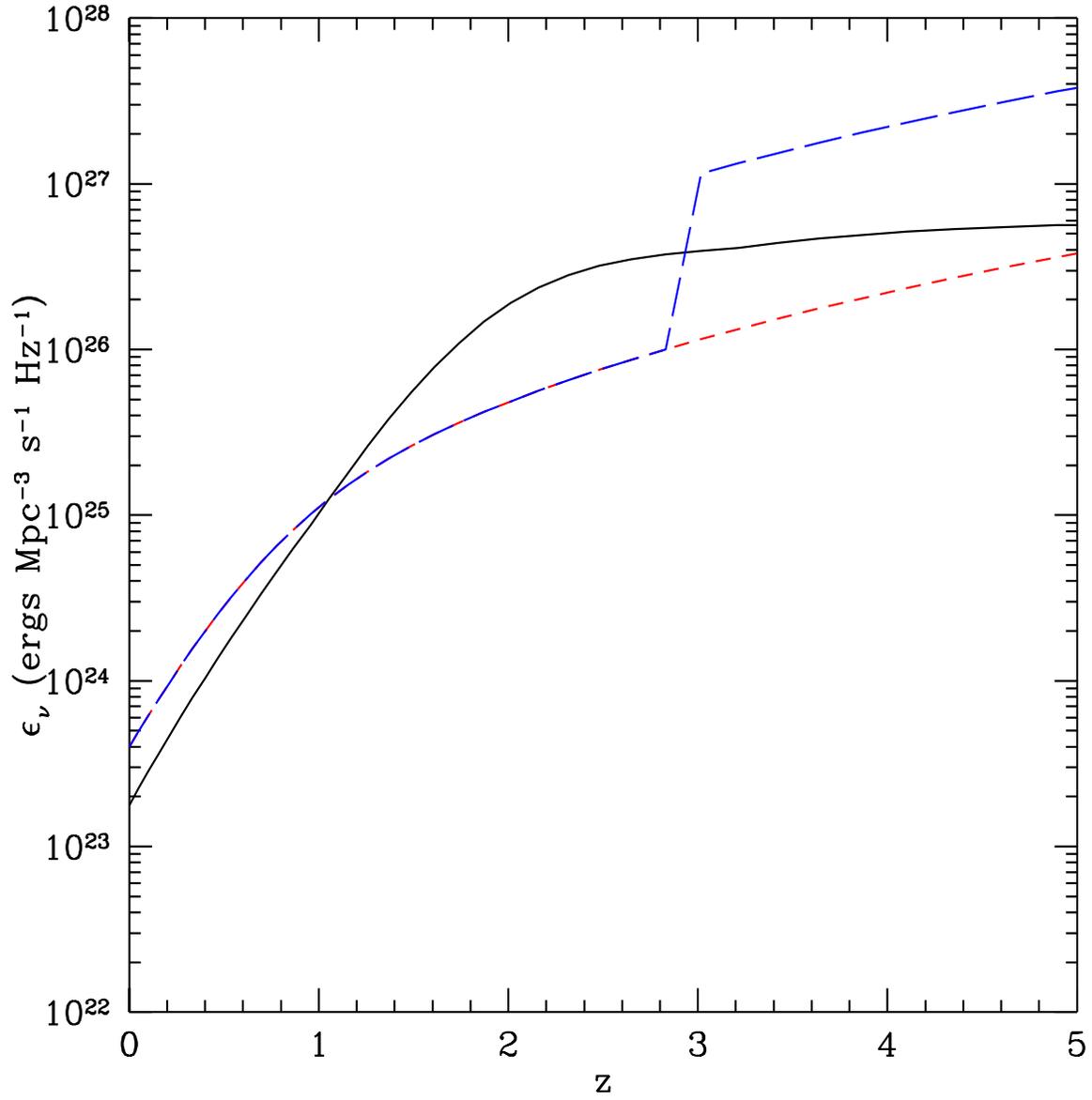,width=1\textwidth,height=0.7\textheight}
\caption{
Specific emissivity at the Lyman limit due to QSOs (black line), YSFGs with low escape fraction (red curve), 
and YSFGs with high escape fraction for $z>3$ (blue line).
}
\end{figure}

\begin{figure}
\psfig{figure=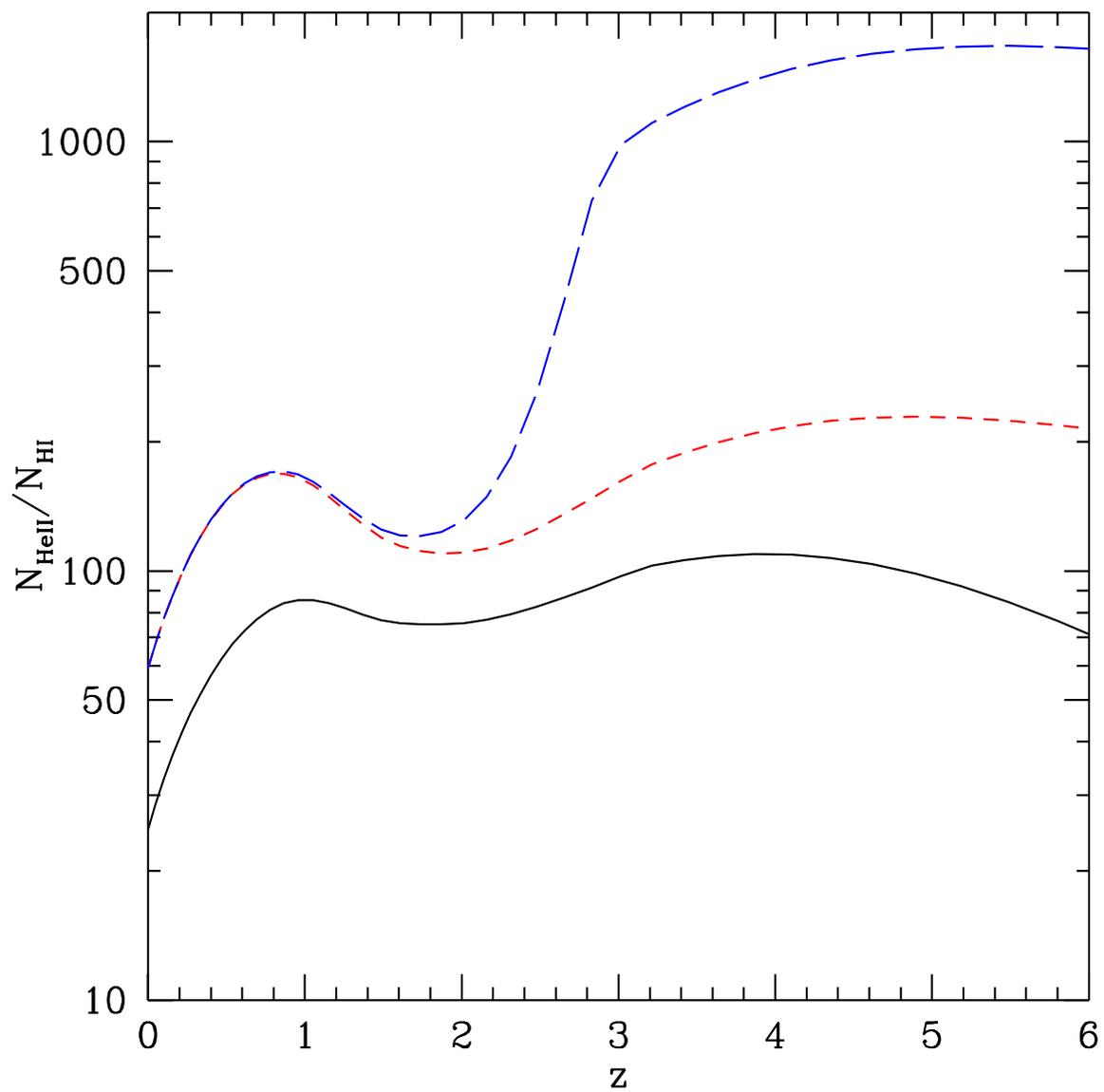,width=1\textwidth,height=0.7\textheight}
\caption{The \HeII/\HI ratio in an IGM in photoionization equilibrium. Colors as in Fig. 7.
}
\end{figure}

\end{document}